\documentclass[journal,12pt,onecolumn]{IEEEtran}

\usepackage[portuguese]{babel}
\usepackage[T1]{fontenc}
\usepackage[utf8]{inputenc}

\usepackage{hyphenat}
\usepackage[table,xcdraw]{xcolor}
\usepackage{natbib}
\setcitestyle{numbers}

\usepackage{geometry}
\usepackage{ae}
\usepackage{amsmath, amsthm, amssymb}
	\newtheorem{assumption}{Suposição}
\usepackage{graphicx}
\usepackage[colorinlistoftodos]{todonotes}

\usepackage{lineno, hyperref}
\modulolinenumbers[5]
\hypersetup{colorlinks={true}, 
						linkcolor={blue},
						filecolor={magenta},      
						urlcolor={cyan},
						allcolors={blue}}
\hypersetup{pdfborder = {0 0 0}}

\begin{document}

\title{Nota técnica: Modelos Implementados}
\author{Coletivo Covid-19BR team}
\maketitle

\begin{abstract}

Esta nota técnica possui o objetivo de realizar uma breve introdução aos modelos de projeção utilizados pelo grupo com a finalidade de projetar cenários futuros para estados e municípios em tempo real, de acordo com o comportamento da doença nos dias prévios. No entanto, os parâmetros podem ser modificados pelo usuário para projetar cenários personalizados. 
O modelo proposto se inicia pelo cálculo do número básico de reprodução para o estado ou município com base na incidência de casos dos últimos 12 dias. Feito isso, é projetada a curva epidemiológica utilizando o modelo epidêmico compartimentado SEIR e, de posse dessa curva, parte dos novos infectados projetados entram em um modelo de simulação para sistemas de saúde em teoria de filas, visando projetar ocupações futuras e colapsos.
\end{abstract}

\section{\textbf{Introdução}}

O crescente avanço da pandemia, declarado pela OMS em março de 2020, causada pelo novo coronavírus (SARS-CoV-2) requer a mobilização de diversos atores para a compreensão da disseminação e os seus impactos em sociedades e sistemas de saúde. \citep{WHOremarks} Em relação ao SARS-CoV-2, a doença decorrente tem apresentado uma alta transmissibilidade e considerável severidade, levando à internação e necessidade de cuidados intensivos em aproximadamente 20\% e 5\% \citep{covid2020severe} dos infectados identificados, respectivamente. Tal comportamento torna a COVID-19 uma doença de alto prejuízo para os indivíduos, sistemas de saúde e sociedade como um todo. 

Contudo, a pandemia de SARS-CoV-2 acontece de forma velada e a capacidade de identificar os casos apenas acontece dias após a infecção, potencializando a transmissão e o espalhamento da doença. Compreender e projetar os padrões de espalhamento (curva epidemiológica) da doença, ao longo do tempo e do espaço geográfico, tornam-se fundamentais para instrumentalizar gestores e lideranças públicas no planejamento de estratégias de enfrentamento e políticas públicas frente ao fenômeno. A modelagem matemática de surtos infecciosos é ferramenta fundamental para elucidação da dinâmica de disseminação do agente e para o impacto de potenciais intervenções. O planejamento para o enfrentamento do surto por parte do poder público e estabelecimentos de saúde depende de evidências que permitam projetar o avanço da doença \citep{WHOreport}. Segundo a organização mundial de saúde, modelos matemáticos ajudaram na compreensão da epidemia de SARS em 2003, e também foram utilizados durante a pandemia de H1N1 em 2009 \citep{WHOreport}.

Esta nota técnica apresenta a metodologia empregada pelo grupo de voluntários \textit{Coletivo Covid-19BR}, composto por professores, pesquisadores e profissionais de mercado, com background em área de saúde e ciências exatas. Esse coletivo surgiu organicamente a partir de iniciativas individuais que foram se somando e construindo em volta da busca de modelos matemáticos que permitissem: (a) analisar/projetar o crescimento da curva epidemiológica da pandemia e, (b) desenvolver um modelo dinâmico de projeção do esgotamento do sistema de saúde para enfrentar a pandemia. Para tanto, utilizamos a metodologia SEIR Bayesiana e a modelagem com base na teoria de filas. Este documento descreve as duas metodologias.

\section{\textbf{Projeção da curva epidemiológica brasileira - Modelo Epidêmico Compartimentado SEIR
}}

\begin{figure}[ht!]
\centering
\includegraphics[width=\textwidth]{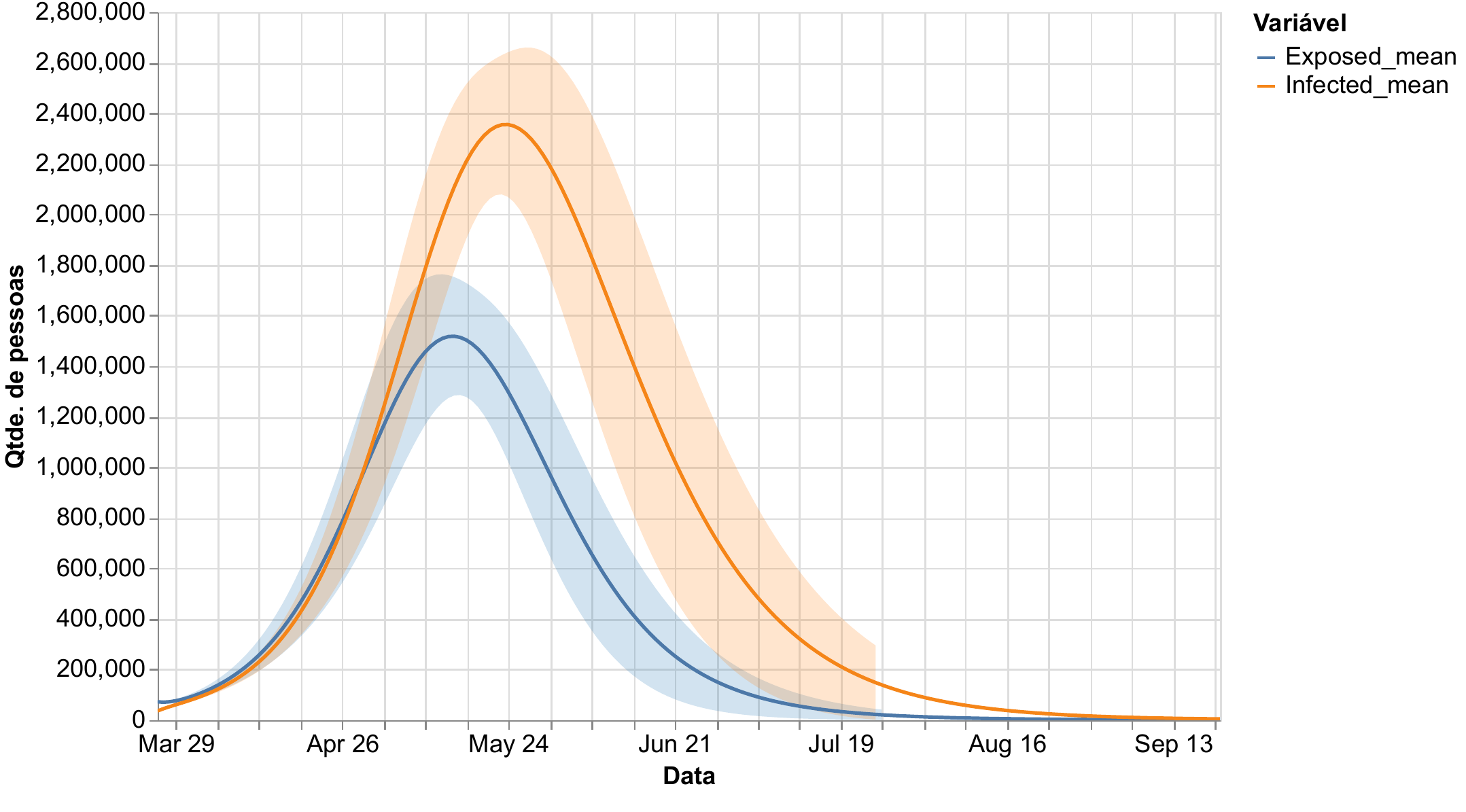}
\caption{Exemplo de curva de projeção do modelo SEIR para casos expostos e infectados.}
\label{fig: ilustração}
\end{figure}

 Modelos compartimentados são uma família de técnicas utilizadas para estudar e prever a dinâmica temporal de contagem e fluxo entre grupos de objetos de estudo \citep{kermack1927contribution}. No caso de doenças infecciosas, são utilizados para prever o alcance e impacto da doença. A população de estudo é particionada em compartimentos disjuntos de forma que os indivíduos em um mesmo compartimento possuam características homogêneas. O saldo total de pessoas em cada grupo é dependente do tempo, pois existe um fluxo de entrada e saída em cada compartimento. A dinâmica do fluxo define o comportamento transiente e estacionário do modelo, e sua intensidade é variável com o tempo; podendo ser desde uma constante até uma função do número total de pessoas por compartimento. Este modelo permite projetar a curva epidemiológica para um dado grupo populacional considerando os parâmetros devidos (Figura \ref{fig: ilustração}).
 
Neste trabalho, foi considerado um modelo com 4 compartimentos: \textbf{S}uscetíveis (população não infectada, porém passível de infecção), \textbf{E}xpostas (infectadas, porém não-contagiosas), \textbf{I}nfectadas (infectadas e contagiosas), e \textbf{R}emovidas (removidas do estudo por óbito ou imunidade); denominado SEIR \citep{kermack1927contribution}.

\subsubsection{Parametrização do modelo}
A evolução temporal (variável $t$) é discretizada a dias. Diariamente, existe uma probabilidade de um indivíduo migrar de compartimento (Tabela \ref{tab:SEIR}). Este valor é variável e depende da contagem de pessoas nos compartimentos $S_t$, $E_t$, $I_t$, e $R_t$; assim como os parâmetros $\alpha^{-1}$ (período médio de incubação), $\gamma^{-1}$ (período médio infeccioso) e $r_0$ (número básico de reprodução).

\begin{table}[h!]
\centering
\caption{Expressões para estimação dos parâmetros do modelo SEIR}
\label{tab:SEIR}
\begin{tabular}{|l|l|l|ll}
\cline{1-3}
\cellcolor[HTML]{000000}{\color[HTML]{FFFFFF} De} & \cellcolor[HTML]{000000}{\color[HTML]{FFFFFF} Para} & \cellcolor[HTML]{000000}{\color[HTML]{FFFFFF} Probabilidade} &  &  \\ \cline{1-3}
Suscetíveis                                       & Expostas                                            & $ p_{SE}(t)=1-exp(\frac{-r_{0}\gamma I_{t}}{N}) $                                     &  &  \\ \cline{1-3}
Expostas                                          & Infectadas                                          & $ p_{EI}(t)=1-exp(-\alpha)$                                       &  &  \\ \cline{1-3}
Infectadas                                        & Removidas                                           & $p_{IR}(t)=1-exp(-\gamma )$                                        &  &  \\ \cline{1-3}
\end{tabular}
\end{table}

Esta componente estocástica resulta em um comportamento não-determinístico para a contagem e fluxo dos compartimentos. Note que $1 - exp(-x)$ é próximo de $x$ para valores de $x$ próximos a zero. Logo, para períodos de incubação e infecção elevados (acima de 1 dia), temos $p_{EI}=\alpha$ e $p_{IR}=\gamma$; que representam as taxas de infecção (razão entre infectados e expostos) e remoção (razão entre removidos e infectados), respectivamente.

Para capturar as incertezas presentes na literatura para os parâmetros $\alpha$ e $\gamma$, foi utilizada uma distribuição LogNormal com 95\% da densidade dentro dos intervalos encontrados na literatura (Tabela \ref{tab:parametrosSEIR}). Já o $r_0$ foi estimado a partir da série histórica de incidência de casos, cujo detalhamento está na seção a seguir. As condições iniciais do modelo $S(0)$, $E(0)$, $I(0)$ e $R(0)$ são consideradas como dadas.

\begin{table}[h!]
\centering
\caption{Valores de referência para os parâmetros X e Y}
\label{tab:parametrosSEIR}
\begin{tabular}{lllll}
\cline{1-4}
\multicolumn{1}{|l|}{\cellcolor[HTML]{000000}{\color[HTML]{FFFFFF} Parâmetro}} & \multicolumn{1}{l|}{\cellcolor[HTML]{000000}{\color[HTML]{FFFFFF} Limite Inferior}} & \multicolumn{1}{l|}{\cellcolor[HTML]{000000}{\color[HTML]{FFFFFF} Limite Superior}} & \multicolumn{1}{l|}{\cellcolor[HTML]{000000}{\color[HTML]{FFFFFF} Referência}} &  \\ \cline{1-4}
\multicolumn{1}{|l|}{$\alpha^{-1}$}                                                      & \multicolumn{1}{l|}{4,1 dias}                                                            & \multicolumn{1}{l|}{7 dias}                                                              & \multicolumn{1}{l|}{\cite{WhoReport2,li2020early}}                                                     &  \\ \cline{1-4}
\multicolumn{1}{|l|}{$ \gamma^{-1}$}                                                      & \multicolumn{1}{l|}{10 dias}                                                             & \multicolumn{1}{l|}{14 dias}                                                             & \multicolumn{1}{l|}{\cite{chen2020clinical}}                                                     &  \\ \cline{1-4}
                                                                               &                                                                                     &                                                                                     &                                                                                & 
\end{tabular}
\end{table}

Formalmente, para os fluxos entre compartimentos, temos
\begin{equation}
SE_{t+1} \sim  Binomial(S_{t},p_{SE}(t)|r_{0},\gamma) 
\label{eq:se}
\end{equation}
\begin{equation}
EI_{t+1} \sim  Binomial(E_{t},p_{EI}(t)|\alpha)   
\label{eq:ei}
\end{equation}
\begin{equation}
IR_{t+1} \sim  Binomial(I_{t},p_{IR}(t)|\gamma)      
\label{eq:ir}
\end{equation}

A partir dos fluxos é possível calcular a contagem

\begin{equation}
S_{t+1} = -SE_{t}
\label{eq:S}
\end{equation}
\begin{equation}
E_{t+1} = SE_{t} - EI_{t}  
\label{eq:E}
\end{equation}
\begin{equation}
I_{t+1} = EI_{t} - IR_{t}  
\label{eq:I}
\end{equation}
\begin{equation}
R_{t+1} = IR_{t}
\label{eq:R}
\end{equation}

\begin{figure}[ht!]
\centering
\includegraphics[width=\textwidth]{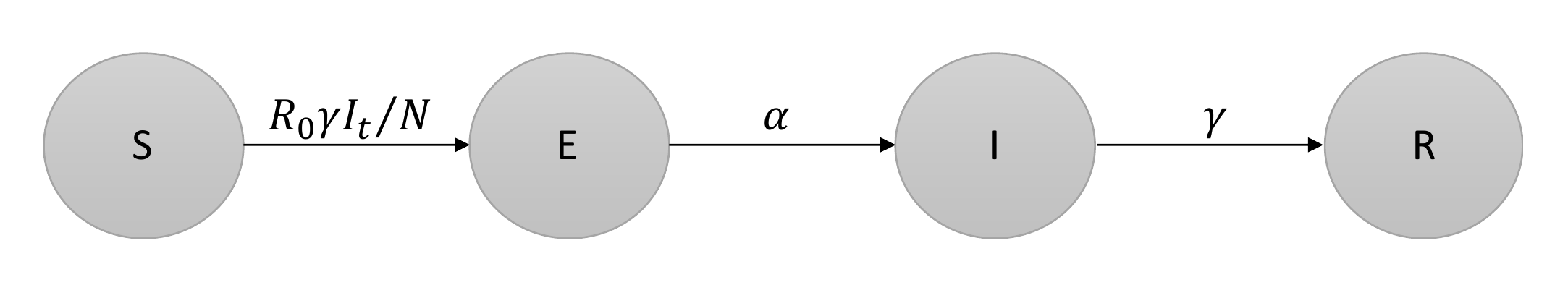}
\caption{Fluxo do modelo SEIR.}
\label{fig: ilustração2}
\end{figure}

\section{\textbf{Estimativa do número de reprodução}}

O número de reprodução foi estimado com base na metodologia descrita em \cite{thompson2019improved,cori2013new}. Assume-se que a incidência observada a cada tempo $t$, $I_t$,  seja descrita adequadamente por uma distribuição de Poisson cujo parâmetro $\lambda_t$ se traduz em uma função do número de reprodução instantâneo, $r_t$, e do potencial infeccioso total, ao longo de todos os indivíduos infectados no tempo $t$, $\Lambda_t$. Formalmente:

\begin{eqnarray}
I&\vee&r_t,\Lambda_t\sim Poisson \left(\lambda_t\right) \\
\label{eq:9}
\lambda_t&=&r_t\cdot\Lambda_t
\end{eqnarray}

Na prática, considera-se $r_t$ constante ao longo de uma janela temporal de tamanho $\tau$, de forma que a equação \ref{eq:9} torna-se $\lambda_t=r_t\cdot\Lambda_{t-\tau}$. O parâmetro $\Lambda_{t-\tau}$ pode então ser interpretado como o potencial infeccioso total ao longo do período $\left[t-\tau,t\right]$, sendo portanto uma função da incidência em cada passo temporal prévio a $I_t$, denotada $I_{t-s}$ para $s\in\left[1,t-\tau\right]$, ponderada pela probabilidade de infecção secundária em cada período de tempo $s$, denotada $w_s$:

\begin{equation}
    \Lambda_{t-\tau}=\sum_{s=1}^{t-\tau}I_{t-s}w_s
\end{equation}

O parâmetro $w_s$ é uma função de massa de probabilidade computada a partir da distribuição estimada para o intervalo serial da doença, isto é, o tempo entre o aparecimento de sintomas em uma pessoa infectada (infector) e o aparecimento de sintomas em pessoas por ela infectadas (infectees). O intervalo serial é comumente usado como proxy para o tempo entre uma infecção primária e as infecções secundárias dela resultantes. Aqui, foi assumido que o intervalo serial é descrito adequadamente por uma distribuição Gamma com média 4,89 e desvio padrão 1,48. Tal parametrização garante 95\% de densidade no intervalo $\left[2,4,8,3\right]$ e foi escolhida com base nas recentes estimativas de intervalo serial para SARS-Cov-2 \citep{du2020serial,Nishiura2020,wu2020nowcasting,zhao2020estimating}.

A inferência sobre $r_t$ se dá via atualização Bayesiana. Sua distribuição a posteriori, condicional aos dados de incidência entre tempo $0$ e tempo $t$ e à função de probabilidade $w_s$, é computada a partir da verossimilhança da incidência observada no intervalo $\left[t-\tau,t\right]$ e da distribuição a priori assumida para $r_t$:

\begin{equation}
    P\left(r_t\vee I_0,I_1,\hdots,I_t,w_s\right)\propto P\left(I_{i\in\left[t-\tau,t\right]}\vee I_{j\in\left[0,t-\tau-1\right]},w_s\right)\cdot P\left(r_t\right)
\end{equation}

Assumindo a priori que $r_t$ segue uma distribuição Gamma, pode-se chegar a uma fórmula fechada para $r_t$ a posteriori haja vista o fato de ser Gamma uma distribuição conjugada à de Poisson, utilizada para computar a verossimilhança. Sendo assim, $P\left(r_t\vee I_0,\hdots,I_t,w_s\right)$ é descrita adequadamente pela distribuição $\Gamma\left(k,\theta\right)$ em que $k$ e $\theta$ são parâmetros de forma e escala, respectivamente, dados por:

\begin{eqnarray}
k&=&k_0+\sum_{j=t-\tau}^tI_j \\
\theta &=&\left(\sum_{j=t-\tau}^t\Lambda_j+\frac{1}{\theta_0}\right)
\end{eqnarray}

\noindent em que $k_0$ e $\theta_0$ são os parâmetros correspondentes da distribuição a priori de $r_t$.

Com base em estimativas anteriores do número básico de reprodução, assumiu-se $k_0=5,12$ e $\theta_0=0,64$, correspondendo a uma média e um desvio padrão de aproximadamente $3,3$ e $1,45$, respectivamente, e posicionando 95\% da densidade a priori entre $1,1$ e $6,7$. Em linhas gerais, essa parametrização abrange as estimativas de 14 estudos anteriores revisados por \cite{liu2020reproductive}. Foi utilizada a janela temporal de $\tau=12$ dias uma vez que janelas menores geraram estimativas instáveis, potencialmente sensíveis a eventual represamento de confirmação de casos diários - em função de falta de testes diagnósticos ou capacidade de processamento destes. Janelas maiores foram descartadas por impossibilitar estimativa para maioria dos estados brasileiros nos quais não haveria dados de incidência suficientes. Estados com dados insuficientes tiveram seu $r_t$ imputado com o valor correspondente ao $r_t$ do Brasil como um todo. O valor de $r_t$ mais recente disponível para cada caso foi utilizado como entrada no modelo SEIR. A metodologia aqui descrita foi implementada em python 3 e reproduz os resultados da função “estimate\_R()”, do pacote de R EpiEstim, utilizando o método “parametric\_si” \citep{Cori2019}.

\section{\textbf{Estimativa da Taxa de Sub-Reportagem}}

Para se estimar a porcentagem dos casos, entre sintomáticos e assintomáticos, que representa o número de confirmações, é utilizada a mortalidade entre os casos confirmados. Para condições de testagem enviesadas, apenas casos mais graves chegam a ser testados, portanto é natural que a mortalidade entre estes seja elevada. Porém a mortalidade entre os já confirmados é, ainda, subestimada pelo fato que alguns virão a óbito. Por isso, apenas dividir o número de mortes pelo número de casos confirmados nos dá um uma razão de fatalidade entre os casos subestimada.

Para chegar a um fator não enviesado, a literatura \citep{nishiura2009early} sugere projetar o número de mortos que teríamos até a data estudada numa situação hipotética em que todos os confirmados morrem, utilizando a distribuição de probabilidade \cite{linton2020incubation} de morte, dado que o paciente morre, entre os dias desde a hospitalização (aqui considerada igual à data de confirmação) até a data atual. A seguinte fórmula expressa o número de mortes acumulado:

\begin{equation}
D_{t} = p_{t}\sum_{i=0}^{t}\sum_{j=0}^{\infty }c_{i-j}f_{i}
\end{equation}

Onde, $D_{t}$ é o número de mortes registradas até o dia t, $c_{t}$ o número de casos confirmados no dia t e $f_{j}$ a probabilidade\cite{linton2020incubation}, dado que morre, de a morte ocorrer após j dias da confirmação e $p_{t}$ é a razão fatalidade-casos não enviesada observada no período. De posse desta taxa, basta utilizar a razão baseline de um estudo abrangente \cite{verity2020estimates} que chegou à razão de 0,657\% de morte entre infectados sintomáticos e assintomáticos.

\begin{equation}
S_{r} = \frac{0,0657}{100*p_{t}}
\end{equation}

Onde $S_r$ é a taxa de sub-reportagem estimada.

\section{\textbf{Projeção para esgotamento do sistema de saúde - Modelo de Filas}}

A teoria de filas busca analisar problemas relacionados a sistemas que envolvam ``linhas de espera''. As bases destes modelos foram iniciadas por \cite{erlang-1908} e depois expandida por \cite{kendall-1951}. Os primeiros sistemas possuíam aspectos determinísticos sendo posteriormente evoluídos para sistemas estocásticos.

\begin{figure}[ht!]
\centering
\includegraphics[width=0.8\textwidth]{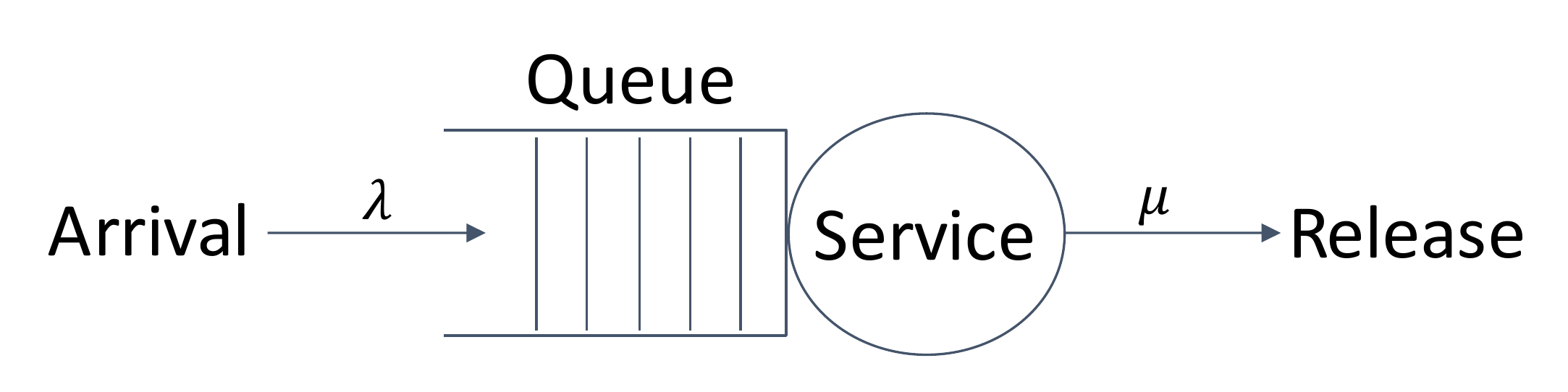}
\caption{Sistema básico de filas.}
\label{fig:figqueue1}
\end{figure}

O problema principal num sistema de filas é resumido conforme o seguinte: sejam $M$ chegadas dentro de um intervalo de tempo ($\lambda$) que utilizam ($s$) servidores por um tempo de serviço ($\mu$), o sistema pode ser representado conforme a Figura \ref{fig:figqueue1}. Estes modelos são baseados num caso especial de cadeias de Markov em tempo contínuo (processo de nascimento e morte).

Os processos de nascimento e morte possuem as seguintes suposições:

\begin{assumption}
Dado um estado do sistema num tempo $t$, $N(t) = n$, a distribuição de probabilidade do tempo remanescente até o próximo nascimento é exponencial com parâmetro $\lambda_n$.
\label{sup:sup1}
\end{assumption}

\begin{assumption}
Dado um estado $N(t) = n$, a distribuição da probabilidade do tempo até a próxima morte é exponencial com parâmetro $\mu_n$.
\label{sup:sup2}
\end{assumption}

\begin{assumption}
Uma única morte ou nascimento pode ocorrer no tempo.
\label{sup:sup3}
\end{assumption}

\section{Aplicações em saúde}

Um dos sistemas de filas mais utilizados em simulações é o \textit{Memoryless/memoryless/s} (M/M/s). Este modelo foi adaptado para o caso de atendimento de hospitais em diversos trabalhos, conforme pode ser analisado em \cite{green-2006, kannapiran-2010, green-2011, poongodi-2013} e \cite{goldwasser-2016}.

As suposições para este sistema são:

\begin{assumption}
O intervalo médio entre chegadas ($\lambda$) é exponencialmente distribuído. O processo de chegada segue distribuição de Poisson.
\label{sup:sup4}
\end{assumption}

\begin{assumption}
O tempo médio de serviço ($\mu$) segue distribuição exponencial.
\label{sup:sup5}
\end{assumption}

\begin{assumption}
Existem $s$ servidores, sendo $s > 1$.
\label{sup:sup6}
\end{assumption}

\begin{assumption}
A taxa de utilização é $\rho = \lambda / s \mu$. O sistema permanece estável enquanto $\lambda < s \mu$ ou $\rho < 1$. Caso contrário, a fila crescerá exponencialmente.
\label{sup:sup7}
\end{assumption}

Por meio deste modelo, torna-se possível obter algumas medidas de eficiência importantes, como o número esperado de indivíduos no sistema ($L = \lambda W$), o tempo de espera em todo o sistema ($W = W_q + 1/ \mu$), o tempo de espera na fila ($W_q = L_q / \lambda$) e o número esperado de indivíduos na fila (expressão \ref{eq:eq_1}).

\begin{equation}
L_q = \frac{P_0 (\lambda /\mu)^2\rho}{s!(1-\rho)^2},
\label{eq:eq_1}
\end{equation}

\noindent
sendo $P_0$ apresentado na expressão \ref{eq:eq_2},

\begin{equation}
P_0 = \left[\sum^{s-1}_{n=0} \frac{(\rho s)^n}{n!} + \frac{\rho^s s^{s+1}}{s! (s-\rho s)}\right]^{-1}.
\label{eq:eq_2}
\end{equation}

\section{\textbf{Modelo proposto para simulação na presença de Covid-19}}

O modelo proposto possui algumas variantes em relação ao sistema M/M/s puro. No presente estudo foi necessário a avaliação de filas de espera por leitos normais e também UTI's, respectivamente os tempos de chegada e serviço diferentes, assim como o número existente de recursos. 

\begin{figure}[ht!]
\centering
\includegraphics[width=0.925\textwidth]{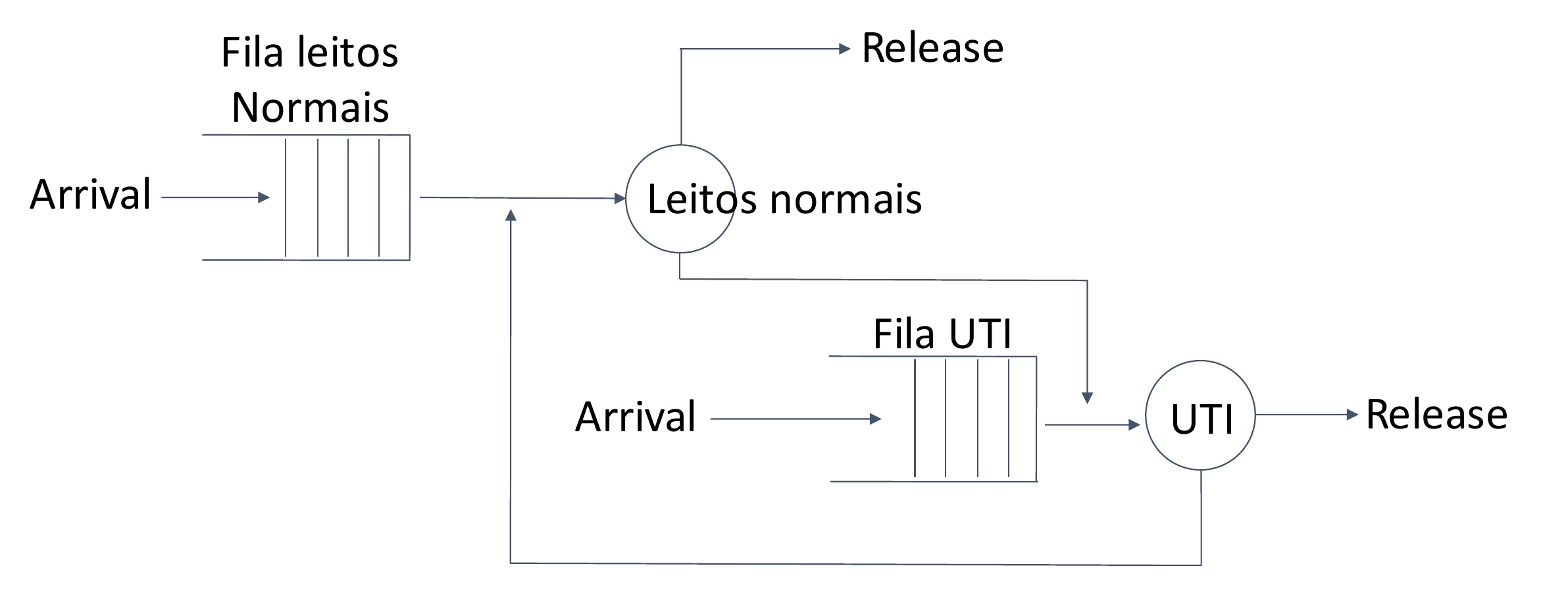}
\caption{Sistema de fila proposto.}
\label{fig:figqueue_sim}
\end{figure}

Também foi necessário a simulação da taxa de chegada num ambiente com Covid-19, neste caso, os valores foram obtidos pelo modelo SEIR (expressões \ref{eq:se} a \ref{eq:R}). Também foi analisada a possibilidade de aumentos de recursos (leitos e UTI's) conforme decisões governamentais. O sistema simulado é apresentado na Figura \ref{fig:figqueue_sim}.

Diante da sofisticação da solução proposta, a obtenção dos parâmetros em uma fórmula fechada se tornou bastante complicada, deste modo, foi realizada uma simulação numérica. Utilizou-se o SimPy (\textit{Discrete event simulation for python}) para a obtenção dos eventos.

\section{\textbf{Origem dos dados}}

Para apoiar os modelos, diversas dados foram obtidos de fontes externas e validados para compor os parâmetros para ambos os simuladores. Na Tabela \ref{tab:tabelaDados} é possível visualizar todas as fontes utilizadas bem como a qual parâmetro elas correspondem.

\begin{table}[]
\caption{Parâmetros e fontes dos dados utilizaods para desenvolvimento da ferramenta.}
\label{tab:tabelaDados}
\centering
\begin{tabular}{|l|l|l|l|l|}
\hline
\textbf{Parâmetro} & \textbf{Valor}/\textbf{Unidade} & \textbf{Fonte}  & \textbf{Ref.}   & \begin{tabular}[c]{@{}l@{}} \textbf{Data} \\ \textbf{extração}\end{tabular} \\ \hline
\begin{tabular}[c]{@{}l@{}}Novos Casos e \\ óbitos\end{tabular} & \begin{tabular}[c]{@{}l@{}}Atualizado \\ diariamente\end{tabular} & Wesley Cota e equipe & \cite{Cota2020}     & \begin{tabular}[c]{@{}l@{}}Atualização\\ diária\end{tabular} \\ \hline
Leitos Normais & & CNES & \cite{BrasilMS} & \begin{tabular}[c]{@{}l@{}}Atualização\\ diária\end{tabular} \\ \hline
Leitos UTI & & CNES  & \cite{BrasilMS} & \begin{tabular}[c]{@{}l@{}}Atualização\\ diária\end{tabular} \\ \hline
Leitos UTI Covid & & CNES & \cite{BrasilMS} & \begin{tabular}[c]{@{}l@{}}Atualização\\ diária\end{tabular} \\ \hline
Novos Leitos & & CNES & \cite{BrasilMS} & \begin{tabular}[c]{@{}l@{}}Atualização\\ diária\end{tabular} \\ \hline
\begin{tabular}[c]{@{}l@{}}Taxa de\\ Hospitalização\end{tabular} & 20,7 - 31,4 & CDC & \cite{covid2020severe} & \begin{tabular}[c]{@{}l@{}}27 de\\ Março 2020\end{tabular} \\ \hline
Taxa de UTI & 4,9 - 11,5 & CDC & \cite{covid2020severe} & \begin{tabular}[c]{@{}l@{}}27 de\\ Março 2020\end{tabular} \\ \hline
Taxa de Óbito & 1,8 - 3,4 & \begin{tabular}[c]{@{}l@{}}CDC e Guan et al\\ (taxas para diferentes\\ fins)\end{tabular} & \cite{covid2020severe}, \cite{guan2020clinical} & \begin{tabular}[c]{@{}l@{}}27 de\\ Março 2020\end{tabular} \\ \hline
População & & IBGE & \cite{IBGE2019} & \begin{tabular}[c]{@{}l@{}}Estimativa\\ IBGE de\\ 01/07/2019\end{tabular} \\ \hline
\begin{tabular}[c]{@{}l@{}}Porcentagem dos\\ infectados que\\ testaram positivo\end{tabular} & \begin{tabular}[c]{@{}l@{}}Calculado para\\ cada lugar\end{tabular} & & \cite{russel2020using}     & \begin{tabular}[c]{@{}l@{}}Atualização\\ diária\end{tabular} \\ \hline
\begin{tabular}[c]{@{}l@{}}Período de\\ infecção\end{tabular} & \begin{tabular}[c]{@{}l@{}}90\% das pessoas\\ zeram a carga\\ detectável de RT-\\ PCR em 10 dias\\ (extrapolação do\\ artigo)\end{tabular} & Liu et al. 2020 & \cite{liu2020viral}     & - \\ \hline
Tempo Incubação & 5 dias & \begin{tabular}[c]{@{}l@{}}OMS (WHO) e Li Q et\\ al\end{tabular} & \cite{WhoReport2}, \cite{li2020early} & - \\ \hline
\begin{tabular}[c]{@{}l@{}}Tempo de estadia\\ médio no leito\\ comum (dias)\end{tabular} & \begin{tabular}[c]{@{}l@{}}8, 9 dias (calculado\\ a partir de dados\\ indiretos do artigo)\end{tabular} & Zhou et al. 2020 & \cite{zhou2020clinical}& - \\ \hline
\begin{tabular}[c]{@{}l@{}}Tempo de estadia\\ médio na UTI\\ (dias)\end{tabular}& 8 dias & Zhou et al. 2020 & \cite{zhou2020clinical} & - \\ \hline
\begin{tabular}[c]{@{}l@{}}Taxa de mortes\\ após estadia na\\ UTI\end{tabular} & $\sim$78\% & Zhou et al. 2020 & \cite{zhou2020clinical}     & - \\ \hline
\begin{tabular}[c]{@{}l@{}}Taxa de pacientes\\ encaminhados\\ para UTI a partir\\ dos leitos\end{tabular} & $\sim$25\% & CDC & \cite{covid2020severe} & - \\ \hline
\end{tabular}
\end{table}

\section{\textbf{Limitações do Modelo}}

Como todo modelo que visa representar a realidade, as projeções apresentadas possuem limitações – as quais estão, a medida do possível, sujeitas a melhorias futuras - que devem ser conhecidas pelo usuário:

\begin{itemize}
    \item Fator básico de reprodução $\left(R_0\right)$: esta métrica é variável com o tempo e depende de medidas governamentais e sociais de combate à pandemia, como isolamento social e quarentena. Tais medidas fogem ao escopo de projeção do nosso modelo, sendo assim é utilizado o fator básico de reprodução para os últimos dias como uma  constante durante o resto da simulação e o resultado da projeção do modelo SEIR, quando utilizado o $\left(R_0\right)$ calculado, deve ser interpretado como relativo ao cenário atual de isolamento, o que pode variar. Além disso, o cálculo do $\left(R_0\right)$ de acordo com os número de casos reportados é sujeito a mudanças na porcentagem de testagem dos infectados. Caso esta última permaneça constante, o número de casos representa o número de infectados multiplicado por uma constante, o que não altera o cálculo do $\left(R_0\right)$. Porém se ao decorrer da pandemia testagem da população variar bruscamente, o $\left(R_0\right)$ estará sujeito a ruídos.
    
    \item O modelo de simulação de filas simplifica todos os leitos do município escolhido como um único sistema de saúde. Porém, na realidade brasileira, existem os sistemas público e privado, que podem colapsar separadamente, assim como os hospitais podem colapsar separadamente.

    \item  De um modo geral, todas as projeções utilizam parâmetros obtidos de estudos estrangeiros relacionados ao comportamento da pandemia em países que tiveram contato com a mesma antes do Brasil. Alguns fatores como os parâmetros com distribuição etária puderam ser ajustados para a pirâmide etária, por exemplo, mas há diversas outras variáveis que podem gerar parâmetros diferentes para a realidade nacional. Sendo assim, parâmetros como letalidade, taxa de hospitalização, tempo médio de estadia nos leitos podem ser diferentes no contexto brasileiro, porém servem como uma boa aproximação. Além disso, existem parâmetros a respeito da doença que ainda não são consenso entre a comunidade científica, como o tempo de duração da doença e o período infeccioso. Vale ressaltar que ter esses parâmetros, ainda que possivelmente enviesados, antecipadamente é uma chance que outros países não tiveram, porém países de pandemia tardia como o Brasil têm.
    
    \item Mais especificamente, a estimativa do $\gamma$ é feita a partir de extrapolações arbitrárias de estudos e resultados observados em testes laboratoriais \cite{duearly} e observações in vivo \cite{he2020temporal} sobre a disseminação da doença, que não visam determinar esse parâmetro. Devido a dificuldade \cite{krylova2013effects} e vieses que todos esses métodos possuem para se estimar ou serem submetidos a extrapolações para determinação do período infeccioso, infere-se que a qualidade da estimativa de tempo de infecciosidade está sujeita a incertezas, de modo que o intervalo de confiança desse período ainda é muito abrangente.
    
\end{itemize}
\newpage

\section{Referências}
\renewcommand{\refname}{\vspace*{-1.5em}}

\bibliographystyle{siam}
\bibliography{tecnotes_references}

\begin{thebibliography}{10}

\bibitem{BrasilMS}
{\sc {Brasil. Ministério da Saúde (MS)}}, {\em Cnes: Cadastro nacional de
  estabelecimentos de saúde. brasília}.
\newblock Disponível em: \url{http://cnes.datasus.gov.br}, 2020.

\bibitem{covid2020severe}
{\sc {CDC COVID-19 Response Team}}, {\em Severe outcomes among patients with
  coronavirus disease 2019 (covid-19)— united states, february 12--march 16,
  2020}, MMWR Morb Mortal Wkly Rep, 69 (2020), pp.~343--346.

\bibitem{chen2020clinical}
{\sc J.~Chen, T.~Qi, L.~Liu, Y.~Ling, Z.~Qian, T.~Li, F.~Li, Q.~Xu, Y.~Zhang,
  S.~Xu, et~al.}, {\em Clinical progression of patients with covid-19 in
  shanghai, china}, Journal of Infection,  (2020).

\bibitem{Cori2019}
{\sc A.~Cori}, {\em Epiestim: Estimate time varying reproduction numbers from
  epidemic curves (version 2.2.1).}
\newblock \url{https://CRAN.R-project.org/package=EpiEstim.}, 2019.

\bibitem{cori2013new}
{\sc A.~Cori, N.~M. Ferguson, C.~Fraser, and S.~Cauchemez}, {\em A new
  framework and software to estimate time-varying reproduction numbers during
  epidemics}, American journal of epidemiology, 178 (2013), pp.~1505--1512.

\bibitem{Cota2020}
{\sc W.~Cota}, {\em Confirmed cases and deaths of covid-19 in brazil}.
\newblock Github \url{https://github.com/wcota/covid19br .}, 2020.
\newblock Acessado em 19/04/2020.

\bibitem{duearly}
{\sc Z.~Du, X.~Xu, Y.~Wu, L.~Wang, B.~J. Cowling, and L.~A. Meyers}, {\em Early
  release-serial interval of covid-19 among publicly reported confirmed cases}.

\bibitem{du2020serial}
\leavevmode\vrule height 2pt depth -1.6pt width 23pt, {\em The serial interval
  of covid-19 from publicly reported confirmed cases}, medRxiv,  (2020).

\bibitem{erlang-1908}
{\sc A.~K. Erlang}, {\em Sandsynlighetsregning og telefonsamtaler}, Nytt
  tidsskrift for Matematik B, 20 (1909).

\bibitem{goldwasser-2016}
{\sc R.~S. Goldwasser, M.~S. Castro~Lobo, E.~F. Arruda, S.~A. Angelo, J.~R.
  Lapa~e Silva, and C.~M. David}, {\em Dificuldades de acesso e estimativas de
  leitos p{\'u}blicos para unidades de terapia intensiva no estado do
  \textsc{R}io de janeiro}, Revista de Sa{\'u}de P{\'u}blica, 50 (2016),
  pp.~1--10.

\bibitem{green-2006}
{\sc L.~Green}, {\em Queueing analysis in healthcare}, Springer, Boston, MA.,
  2006, pp.~281--307.

\bibitem{green-2011}
\leavevmode\vrule height 2pt depth -1.6pt width 23pt, {\em Handbook of
  Healthcare Delivery Systems}, Taylor \& Francis, London, 2011.

\bibitem{guan2020clinical}
{\sc W.-j. Guan, Z.-y. Ni, Y.~Hu, W.-h. Liang, C.-q. Ou, J.-x. He, L.~Liu,
  H.~Shan, C.-l. Lei, D.~S. Hui, et~al.}, {\em Clinical characteristics of
  coronavirus disease 2019 in china}, New England Journal of Medicine,  (2020).

\bibitem{he2020temporal}
{\sc X.~He, E.~H. Lau, P.~Wu, X.~Deng, J.~Wang, X.~Hao, Y.~C. Lau, J.~Y. Wong,
  Y.~Guan, X.~Tan, et~al.}, {\em Temporal dynamics in viral shedding and
  transmissibility of covid-19}, Nature Medicine,  (2020), pp.~1--4.

\bibitem{IBGE2019}
{\sc {IBGE (Instituto Brasileiro de Geografia e Estatística}}, {\em Tabelas de
  estimativas para 1º de julho de 2019, atualizadas e enviadas ao tcu após a
  publicação no dou. 2019 jun.}
\newblock Disponível em:
  \url{https://www.ibge.gov.br/estatisticas/sociais/populacao/9103-estimativas-de-populacao.html},
  2019.

\bibitem{kannapiran-2010}
{\sc R.~P. Kannapiran and K.~L. Teow}, {\em Queueing for healthcare}, Journal
  of Medical Systems, 36 (2010), pp.~541--54.

\bibitem{kendall-1951}
{\sc D.~G. Kendall}, {\em Some problems in the theory of queues}, Journal of
  Royal Statistics Society, 13 (1951), pp.~151--173.

\bibitem{kermack1927contribution}
{\sc W.~O. Kermack and A.~G. McKendrick}, {\em A contribution to the
  mathematical theory of epidemics}, Proceedings of the royal society of
  london. Series A, Containing papers of a mathematical and physical character,
  115 (1927), pp.~700--721.

\bibitem{krylova2013effects}
{\sc O.~Krylova and D.~J. Earn}, {\em Effects of the infectious period
  distribution on predicted transitions in childhood disease dynamics}, Journal
  of The Royal Society Interface, 10 (2013), p.~20130098.

\bibitem{li2020early}
{\sc Q.~Li, X.~Guan, P.~Wu, X.~Wang, L.~Zhou, Y.~Tong, R.~Ren, K.~S. Leung,
  E.~H. Lau, J.~Y. Wong, et~al.}, {\em Early transmission dynamics in wuhan,
  china, of novel coronavirus--infected pneumonia}, New England Journal of
  Medicine,  (2020).

\bibitem{linton2020incubation}
{\sc N.~M. Linton, T.~Kobayashi, Y.~Yang, K.~Hayashi, A.~R. Akhmetzhanov, S.-m.
  Jung, B.~Yuan, R.~Kinoshita, and H.~Nishiura}, {\em Incubation period and
  other epidemiological characteristics of 2019 novel coronavirus infections
  with right truncation: a statistical analysis of publicly available case
  data}, Journal of clinical medicine, 9 (2020), p.~538.

\bibitem{liu2020reproductive}
{\sc Y.~Liu, A.~A. Gayle, A.~Wilder-Smith, and J.~Rockl{\"o}v}, {\em The
  reproductive number of covid-19 is higher compared to sars coronavirus},
  Journal of travel medicine,  (2020).

\bibitem{liu2020viral}
{\sc Y.~Liu, L.-M. Yan, L.~Wan, T.-X. Xiang, A.~Le, J.-M. Liu, M.~Peiris, L.~L.
  Poon, and W.~Zhang}, {\em Viral dynamics in mild and severe cases of
  covid-19}, The Lancet Infectious Diseases,  (2020).

\bibitem{nishiura2009early}
{\sc H.~Nishiura, D.~Klinkenberg, M.~Roberts, and J.~A. Heesterbeek}, {\em
  Early epidemiological assessment of the virulence of emerging infectious
  diseases: a case study of an influenza pandemic}, PLoS One, 4 (2009).

\bibitem{Nishiura2020}
{\sc H.~Nishiura, N.~M. Linton, and A.~R. Akhmetzhanov}, {\em Serial interval
  of novel coronavirus (2019-ncov) infections}, medRxiv,  (2020).

\bibitem{poongodi-2013}
{\sc T.~Poongodi and S.~Muthulakshmi}, {\em Control chart for waiting time in
  system of (m/m/1):(infinite/fcfs) queuing model}, International Journal of
  Computer Applications, 63 (2013), pp.~48--53.

\bibitem{russel2020using}
{\sc T.~Russel, J.~Hellewell, S.~Abbot, et~al.}, {\em Using a delay-adjusted
  case fatality ratio to estimate under-reporting}, Available at the Centre for
  Mathematical Modelling of Infectious Diseases Repository, here,  (2020).

\bibitem{thompson2019improved}
{\sc R.~Thompson, J.~Stockwin, R.~van Gaalen, J.~Polonsky, Z.~Kamvar,
  P.~Demarsh, E.~Dahlqwist, S.~Li, E.~Miguel, T.~Jombart, et~al.}, {\em
  Improved inference of time-varying reproduction numbers during infectious
  disease outbreaks}, Epidemics, 29 (2019), p.~100356.

\bibitem{verity2020estimates}
{\sc R.~Verity, L.~C. Okell, I.~Dorigatti, P.~Winskill, C.~Whittaker, N.~Imai,
  G.~Cuomo-Dannenburg, H.~Thompson, P.~G. Walker, H.~Fu, et~al.}, {\em
  Estimates of the severity of coronavirus disease 2019: a model-based
  analysis}, The Lancet infectious diseases,  (2020).

\bibitem{WhoReport2}
{\sc {WHO}}, {\em Report of the who-china joint mission on coronavirus disease
  2019 (covid-19).}, 2020.

\bibitem{WHOreport}
{\sc WHO}, {\em Weekly epidemiological record}.
\newblock \url{https://www.who.int/wer/2009/wer8434.pdf}, 2020.

\bibitem{WHOremarks}
\leavevmode\vrule height 2pt depth -1.6pt width 23pt, {\em Who
  director-general's opening remarks at the media briefing on covid-19 - 11
  march 2020}.
\newblock
  \url{https://www.who.int/dg/speeches/detail/who-director-general-s-opening-remarks-at-the-media-briefing-on-covid-19---11-march-2020},
  2020.

\bibitem{wu2020nowcasting}
{\sc J.~T. Wu, K.~Leung, and G.~M. Leung}, {\em Nowcasting and forecasting the
  potential domestic and international spread of the 2019-ncov outbreak
  originating in wuhan, china: a modelling study}, The Lancet, 395 (2020),
  pp.~689--697.

\bibitem{zhao2020estimating}
{\sc S.~Zhao, D.~Gao, Z.~Zhuang, M.~Chong, Y.~Cai, J.~Ran, P.~Cao, K.~Wang,
  Y.~Lou, W.~Wang, et~al.}, {\em Estimating the serial interval of the novel
  coronavirus disease (covid-19): A statistical analysis using the public data
  in hong kong from january 16 to february 15, 2020}, medRxiv,  (2020).

\bibitem{zhou2020clinical}
{\sc F.~Zhou, T.~Yu, R.~Du, G.~Fan, Y.~Liu, Z.~Liu, J.~Xiang, Y.~Wang, B.~Song,
  X.~Gu, et~al.}, {\em Clinical course and risk factors for mortality of adult
  inpatients with covid-19 in wuhan, china: a retrospective cohort study}, The
  Lancet,  (2020).

\end{thebibliography}

\end{document}